\documentclass[twocolumn,showpacs,prbprintnumbers,prb,fleqn,superscriptaddress]{revtex4}
\usepackage{epsfig}
\usepackage{graphicx}
\usepackage{amsfonts}
\usepackage[dvips]{color}
\usepackage{color}

\newcommand{\ct}{\cite}
\newcommand{\la}{\lambda}
\newcommand{\bi}{\bibitem}
\newcommand{\be}{\begin{equation}}
\newcommand{\ee}{\end{equation}}
\newcommand{\ba}{\begin{eqnarray}}
\newcommand{\ea}{\end{eqnarray}}
\newcommand{\al}{\alpha}

\newcommand{\ga}{\gamma}

\newcommand{\ket}[1]{|#1\rangle}
\newcommand{\de}{\delta}

\begin{document}

\title{ The scaling of the decoherence factor of a qubit coupled to a spin chain driven
across quantum critical points}
\author{Tanay Nag}
\affiliation{Department of Physics, Indian Institute of Technology Kanpur,
Kanpur 208 016, India}
\author{Uma Divakaran}
\affiliation{Department of Physics, Indian Institute of Technology Kanpur,
Kanpur 208 016, India}
\author{Amit Dutta}
\affiliation{Department of Physics, Indian Institute of Technology Kanpur,
Kanpur 208 016, India}

\begin{abstract}
 
We study the scaling of the decoherence factor of a qubit (spin$-1/2$) using the central spin model in which the central spin (qubit) is globally coupled to a transverse
XY spin chain. The aim here is to study the non-equilibrium generation of decoherence
when the spin chain is driven across (along) quantum critical points (lines) and derive the scaling
of the decoherence factor in terms of the driving rate and some of the exponents associated with the quantum critical points. Our studies show that  the scaling
of logarithm of decoherence factor is identical to that of the defect density in the final state of the spin chain following a quench across 
isolated quantum critical points for both linear and non-linear 
variations of a parameter even if the defect density may not satisfy the standard Kibble-Zurek scaling. However, one finds an interesting deviation when
the spin chain is driven along a critical line.  Our analytical predictions 
are in complete agreement with numerical results.  Our study, though limited to  integrable two-level systems, points to the existence of a
 universality in the scaling of the decoherence factor  which is not necessarily identical to
the scaling  of the defect density.
\end{abstract}

\maketitle

When  a quantum many-body system is slowly driven across a quantum critical point (QCP) \cite{sachdev99}
by varying a parameter in the Hamiltonian, 
defects are generated in the final state; this is a consequence of the diverging relaxation time 
close to the QCP, so that 
the dynamics
is no longer adiabatic however slow may the variation be \ct{zurek96,zurek05,polkovnikov05}.  If a parameter $\la$ of the Hamiltonian describing a $d-$dimensional system is changed linearly 
as $\la(t)=t/\tau,~~-\infty<t<\infty$, (with the QCP at $\la=0$), the defect density ($n$) in the final state  satisfies the Kibble-Zurek (KZ) scaling relation,
\cite{zurek96,zurek05,polkovnikov05,damski05,mukherjee07,polkovnikov11,dutta10}
$n \sim \tau^{-\nu d/(\nu z +1)}$; here, $\tau$ is the inverse rate of quenching, 
and $\nu$ and $z$ are the correlation length and dynamical critical exponents, respectively, associated with the QCP.

In parallel, there are a plethora of studies which connect quantum information theory
to quantum critical systems (for a review, see \ct{polkovnikov11,dutta10}). One of the major issues in this regard is  
 the study of decoherence,  namely, the loss of coherence in a quantum system due to its interaction with the environment \ct{zurek03}. 
To elucidate these studies, the central spin model (CSM) 
has been proposed \cite{quan06}.
In this model, a central spin (CS) (i.e., the qubit) has a global interaction with a quantum many body system (e.g., with all the spins of a quantum spin chain)
 which acts as the environment. The interaction between the qubit and the environment in fact provides two channels of time evolution of the environmental spin chain.
It has been observed that the purity of the CS is  given in terms of the Loschmidt echo (LE) or the 
decoherence factor (DF) which is the measure of the square of the overlap of the wave function evolved along the two different channels as a function of time. 
The LE or DF which appears in the off-diagonal term of
the reduced density matrix of the qubit  is minimum
at the QCP signifying a maximum loss of coherence close to it
which can be used as an indicator of quantum criticality \ct{quan06,venuti10,cucchietti07}.

At this point, 
the natural question would be what happens when the environment
 is driven following some protocol across a QCP.
In a recent work, Damski $et~al$ \cite{damski11}, studied the
decoherence of the CS by coupling it to a transverse Ising spin chain which
is driven across the QCP by a linear variation of  the transverse field  and showed that in the limit of weak coupling,   
the logarithm of the non-adiabatic part of the DF (arising due to the contribution of the low-energy modes
close to the critical mode and denoted by $D_{non-ad}$)
 satisfies an identical scaling  to that of
$n$, given by  $\ln D_{non-ad} \sim \tau^{-1/2}$. 


In present work, we consider a version of the CSM in which environmental spin chain
is chosen to be an anisotropic $XY$ spin chain which has a rich phase diagram and thereby enables us to study 
the scaling of the DF when the environment is quenched
across different critical and multicritical points as well as gapless critical lines.
 The 
spin chain is exactly solvable using Jordan-Wigner(JW)
transformations \ct{lieb61} and is reducible to a decoupled two-level problem in the Fourier space and hence
the Schr\"odinger equations describing the evolution of these two levels can be analytically
solved for all times.  We shall, however, emphasize on an alternative method introduced in \ct{pollmann10}.
This method, valid away from the QCP, exploits
 the two-level nature of the reduced Hamiltonian and the exact expression for the 
probability of non-adiabatic transition at the final time
as given by the Landau-Zener (LZ) transition formula \ct{landau}. Both the results
however lead to identical scaling relations which are also verified numerically.
 Moreover, this alternative 
approach also allows us to calculate the scaling of the DF for a non-linear variation of the  quenching parameter  though the LZ formula
is not known exactly. 

Let us emphasize that our focus here is limited to only low-energy modes
close to the critical mode for which the energy gap vanishes at the QCP. The high energy modes
on the other hand, evolve adiabatically throughout the dynamics; though these modes contribute to the dynamics of decoherence non-trivially
through the fidelity factor, they do not alter the scaling relation of the DF \cite{damski11}.

Let us first clarify the connection between the scaling of the DF
and $n$, that  we are interested in. We shall assume weak coupling between
the qubit and the environment  and work within the appropriate range of time; under these circumstances, 
for all the quenching schemes  discussed  here  (achieved by changing
a parameter $\la=t/\tau$), we find   the scaling relations: (i) $\ln D_{non-ad}\sim (-t^2 f(\tau))$, if QCP is at $\la=0$ and
(ii) $\ln D_{non-ad} \sim \{-(t-\la_0 \tau)^2 f(\tau)\}$, if the QCP is at $\la_0$ (as happens for quenching through a MCP discussed below). We explore the scaling
of this function $f(\tau)$ (which is found to be linear in the size of the spin chain and quadratic in system-environment coupling) with  $\tau$ and address the question whether that is identical to the scaling
 of $n$. 
However, to eliminate $t$, one could further substitute $t =\la \tau$, to obtain the scaling
$ \ln D_{non-ad} \sim -\la^2 \tau^2 f(\tau)$ (or $\sim - (\la -\la_0)^2\tau^2 f(\tau)$ for case (ii)), but it should be emphasized that the non-trivial
scaling of  $\ln D_{non-ad}$ with $\tau$ is provided by that of $f(\tau)$.
Our studies reveal that in the cases when the environment
is driven through an isolated QCP
or a multicritical point (MCP), the scaling of $\ln D_{non-ad}$ (or precisely that of  $f(\tau)$)   is the same as that  of the defect density
in the final state following a quench for both linear and non-linear quenches. However, there are situations when this generic
connection do not hold. For example, when the environment
is driven 
along a critical line across the MCP, we arrive at a scaling which is significantly different from that
of  $n$.

\begin{figure}
\includegraphics[height=2.0in]{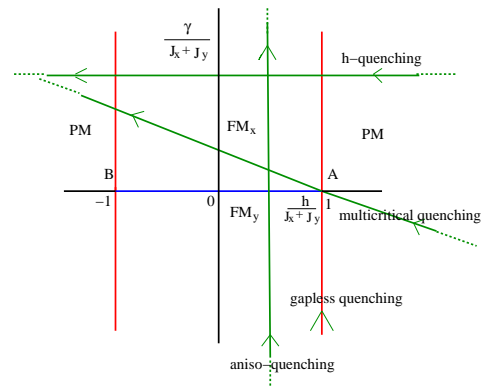}
\caption{(Color Online) The phase diagram of Hamiltonian (\ref{eq_xy1})
in the $h/(J_x+J_y)-\ga/(J_x+J_y)$ plane.
The two red vertical lines are the transverse Ising critical lines between the
ferromagnetic (FM) and paramagnetic (PM) phases. The blue horizontal 
line for $-1<h<1$ corresponds to the anisotropic critical line
separating the two ferromagnetic phases FM$_{\rm x}$ and FM$_{\rm y}$, with ordering in $x$ and $y$ directions, respectively.
Points denoted by A and B represent the two multicritical points. 
 {Green lines represent different quenching paths discussed in this paper}.}
\label{fig_xy}
\end{figure}

The Hamiltonian $H_E$ of the environment is the $XY$
spin chain in a transverse field consisting of $N$ spins given by \ct{bunder99}
\be
H_E=-\sum_{i=1}^N [J_x \sigma_i^x\sigma_{i+1}^x +J_y \sigma_i^y\sigma_{i+1}^y
+h \sigma_i^z] 
\label{eq_xy1}
\ee
which is coupled to the spin-1/2 qubit by a Hamiltonian $H_{SE}$.
In the following, we shall define the anisotropy parameter $\gamma=J_x-J_y$, and the parameter $(J_x+J_y)$ will be set equal to unity in
all cases except for the quenching through a MCP.
The model is exactly solvable by  JW transformation \ct{bunder99};  the phase diagram is shown in Fig \ref{fig_xy}.

Let us introduce the notion of  DF by considering the situation in which the transverse field $h$ 
is quenched as $h(t)=1-t/\tau$, and the qubit is coupled to the 
time dependent transverse field of (\ref{eq_xy1}) through the Hamiltonian
$H_{SE}=-\delta\sum_{i=1}^N\sigma_i^z\sigma_S^z$,
where $\sigma_i^z$ is the $i-$th spin of the $XY$ chain and $\sigma_S^z$ represents that of the
qubit. The system crosses the Ising critical points at $h= 1$ and $h=-1$ with critical mode 
$k_c$ given by $k_c=\pi$, and $0$, respectively.
We choose
the qubit to be initially (at $t \to -\infty$)  in a pure state superposition 
 $|\phi_S(t\to-\infty)\rangle= c_1|\uparrow\rangle+c_2|\downarrow\rangle$, where $|\uparrow\rangle$ and 
$|\downarrow \rangle$ represent up and down states of the CS, respectively, and the
environment is in the ground state $|\phi_E (t\to -\infty)\rangle = |\phi_g \rangle$.  The
ground state of the composite Hamiltonian $H_E + H_{SE}$, at $t \to -\infty$, is given by the direct product
$|\psi(t \to -\infty)\rangle=|\phi_S(t \to -\infty)\rangle \otimes |\phi_g\rangle$. 
It can be shown that at a later  time $t$, the composite wave function is given by
$|\psi(t)\rangle 
=c_1|\uparrow\rangle \otimes |\phi_+\rangle + c_2|\downarrow \rangle \otimes |\phi_-\rangle$,
where $|\phi_{\pm}\rangle$ are the wavefunctions evolving with the environment
Hamiltonian $H_E(h\pm\delta)$
given by the  Schr\"odinger equation 
$i {\partial}/{\partial t}|\phi_{\pm}\rangle=\hat H[h(t)\pm\delta]|\phi_{\pm}\rangle.$
We therefore find that the coupling $\de$ essentially provides two channels of evolution of the environmental
wave function
with the transverse field $h+\de$ and $h-\de$, respectively. 

It is straightforward to show that the decoherence factor $D(t)$
defined as $|\langle \phi_+(t)|\phi_-(t)\rangle|^2$, is the off-diagonal element of the reduced density matrix of the qubit \ct{damski11}.
To evaluate $D(t)$, we rewrite the Hamiltonian (\ref{eq_xy1})
with modified $h$ (due to the coupling $\de$)
in terms of JW fermions which then  can be decoupled into a sum of
independent $(2 \times 2)$ Hamiltonians in the Fourier space \cite{lieb61,bunder99}.
In the basis $\ket{0}$ and $\ket{k,-k}$, which represent no quasiparticle, and  quasiparticles with momentum $k$ and $-k$,
respectively, the Hamiltonian $H_E$ can be written as
\begin{eqnarray}
H_E^\pm(t)&=&\sum_k H_k^\pm(t),~~{\rm where,} \nonumber \\
H_k^\pm (t)&=& 2 
 \left(
 \begin{array}{cc}
    h(t) \pm \delta + \cos k & \ga \sin k  \\
\ga \sin k & -(h(t) \pm \delta + \cos k)   \\
 \end{array}
 \right).
 \label{eq_math}
 \nonumber
\end{eqnarray}
The general wave function for $H_E$ at any instant $t$ can be written as  
\begin{eqnarray}
|\phi^{\pm}(t)\rangle&=&\prod_k |\phi_k^\pm(t)\rangle = \prod_{k>0} \left[u_k^\pm(t) |0\rangle + v_k^\pm(t)|k,-k\rangle \right].\nonumber
\end{eqnarray}
The coefficients $u_k^\pm$ and $v_k^\pm$ are obtained by solving the Schr\"odinger
equation 
$i{\partial}/{\partial t} \left(u_k^\pm(t),v_k^\pm(t)\right)^{T}= H_k^\pm(t) \left(u_k^\pm(t),v_k^\pm(t)\right)^T$
where $A^T$ represents the transpose operation of the row matrix $A$.
Hence, the expression of $D(t)$ is given by 
$\prod_k F_k(t)=  \prod_k |\langle {\phi_k(h(t)+\delta)}|{\phi_k(h(t)-\delta)}\rangle|^2$, or,
\begin{eqnarray}
D(t)=\exp\left[
\frac{N}{2\pi} \int_0^{\pi} dk~\ln F_k
\right]
\label{eq_gendecoh}
\end{eqnarray}
where $F_k$ can be written in terms of $u_k^\pm$ and $v_k^\pm$. 
We reiterate that we shall focus in the limit of small $\de$ and consider only the low-energy modes
%
which  show non-adiabatic behavior close to the QCP.  On the other hand, the high energy modes
evolve adiabatically  and their overlap is close to unity.  This method can be useful for
exact solution as well as numerical estimation of $D(t)$.

We shall however introduce a simpler method for analytical calculations
that exploits the $(2 \times 2)$ nature of the reduced Hamiltonian to calculate $F_k(t)$ \ct{pollmann10}. Far away
from the QCP ($|h(t)| {\gg} 1 $ ($t \to +\infty$)) i.e., after crossing both the QCPs, we can write 
$\ket{\phi_k(h+\delta)}=u_k\ket 0 + v_k e^{-i\Delta^+ t}\ket{k,-k}$,
and $\ket{\phi_k(h-\delta)}=u_k\ket 0+e^{-i\Delta^- t}v_k\ket{k,-k}$ where $\Delta^+=4\sqrt{(h+\delta+1)^2 + \gamma^2 \sin k^2}$ 
and $\Delta^-=4\sqrt{(h-\delta+1)^2 + \gamma^2 \sin k^2}$ are the
energy difference between the states $\ket{0}$ and $\ket{k,-k}$ when the 
transverse field is equal to $h+\delta$ and $h-\delta$, respectively. 
In writing the above expression, we make use of the fact that excitations occur only in
the vicinity of QCPs. Following that the wavefunctions ($|\phi^{\pm}(t)\rangle$) evolve adiabatically picking up  the appropriate phase factor with time.
 At the same time, the coefficients $u_k$ and $v_k$ can
be found to be
$|u_k|^2=1-p_k$ and $|v_k|^2=p_k$ where $p_k$ is the Landau-Zener probability
of excitations for the mode $k$ given by $p_k=\exp(-2\pi \tau \gamma^2 \sin^2 k)$ \cite{landau}.
Combining all these, we find 
\begin{eqnarray}
F_k(t)&=&|\langle\phi_k(h(t)+\delta)|\phi_k(h(t)-\delta)\rangle|^2\nonumber \\
&=&\left| |u_k|^2+|v_k|^2e^{-i(\Delta^+-\Delta^-) t}\right |^2
\label{eq_dt1},
\end{eqnarray}
which can be recast in the vicinity of  the quantum critical point at $h=1$ to the form 
$\Delta=(\Delta^+-\Delta^-)/2$,
\begin{eqnarray}
F_k(t)&=&1-4p_k(1-p_k) \sin^2(\Delta t )\nonumber\\
&=&1-4\left[e^{-2\pi \tau \gamma^2 k'^2 } - e^{-4\pi \tau \gamma^2 k'^2} \right] \sin^2 (4\delta t)
\label{eq_fknonad}
\end{eqnarray}
where $\sin k$ has been expanded near the critical modes $k= \pi$, with $k'=\pi-k$  and we have taken the limit $\de\to 0$.
The above expression is identical to that given in  \cite{damski11}  derived via the exact solution
of the Schr\"odinger equation. 

The DF  is the product of the contribution from the modes evolving adiabatically (given by fidelity)
and the modes evolving non-adiabatically denoted  by$D_{non-ad}$.
The expression 
of $D_{non-ad}(t)$ 
due to the non-adiabatic dynamics of  modes $k \simeq \pi$
after crossing the critical point $h=1$ can be obtained from Eq.~(\ref{eq_gendecoh}) 
in the following way:
in the limit $\de \to 0$ (or more precisely $(\delta t) \to 0$),
one can approximate $\sin^2 4\delta t \approx 16 \delta^2 t^2$ which results to
\begin{eqnarray}
D_{non-ad} (t)&=&\exp\frac {N}{2\pi} \int_0^{\infty} dk \nonumber\\
&\ln& \left[
1-\left(e^{-2\pi \tau \gamma^2 k'^2 } - e^{-4\pi \tau \gamma^2 k'^2} \right)
 64 \delta^2 t^2\right]
\end{eqnarray}
where we have extended the limit of integration  to $\infty$ since only the modes close to 
the critical modes contribute in the limit of large $\tau$. Using the fact that $\ln (1-x) \sim -x$, for small $x$,
it can be further shown that $D_{non-ad}$ is given by
\begin{eqnarray}
D_{non-ad}(t) \sim \exp \{ -8(\sqrt{2}-1) N \delta^2 t^2/(\gamma \pi \sqrt \tau)\}.
\label{eq_expdecay}
\end{eqnarray}
 
It is worth noting  that the periodicity in time as in Eq.~(\ref{eq_fknonad}) is lost and there is an exponential decay as
shown in Eq.~(\ref{eq_expdecay}). This Gaussian form holds true when $t \ll 1/\de$ and $\de\to 0$; clearly the time range over which this
is applicable increases with decreasing $\de$. Otherwise, a sinusoidal variation is observed.
A similar expression can be obtained
for the low $k$ modes excited after crossing the $h=-1$ critical point.
We find that $\ln D_{non-ad}(t)$ rather $f(\tau) =  \{8(\sqrt{2}-1) N \delta^2\}/(\pi \gamma \sqrt \tau)$ varies as $1/\sqrt{\tau}$, a scaling  which is identical
to that of $n$ (with $d=\nu=z=1$).

\begin{figure}
\includegraphics[height=1.3in]{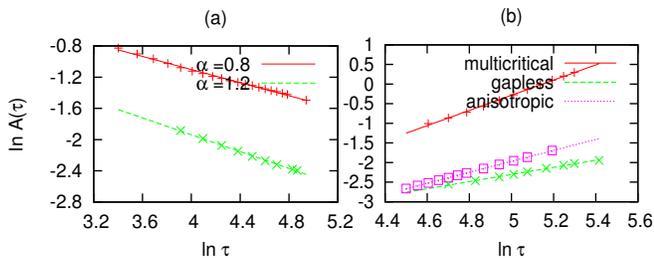}
\caption{(Color Online)(a) The variation of $\ln A(\tau)$ with $\ln \tau$ 
for two different values of $\alpha$ 
with $N=300$ and $\delta=0.0001$: (i) $\alpha=0.8$ (red line) for fixed $t=1500$ with slope $-0.42$ and,
(ii) $\alpha=1.2$ (green dashed line) for $t=670$ with slope $-0.54$;
both the slope values are very close to the predicted ${-\alpha/(\alpha+1)}$.
(b) The same 
for three different types of qubit-environment coupling with $\delta=0.0001$ and $\al=1$.
(i) Line (red) corresponds to quenching across the multicritical point
A in Fig. \ref{fig_xy} by varying $J_x \sim t/\tau$ 
with $J_x=7$ and $N=500$;
the slope $\simeq 1.9$ 
(ii) Dotted line (pink) corresponds to quenching $\gamma \sim t/\tau$ at $h=0.5$
(across the anisotropic QCP)
with $\gamma=6.5$ and 
N=200, and the slope  $\simeq 1.4$
(iii)Dashed line (green) is for $\gamma$ quenching as $t/\tau$ along the
gapless line $h=1$ with slope $\simeq 0.9$,
for $\gamma=5$
and N=400.  Analytically predicted values are $11/6$, $3/2$ and $1$, respectively.
}
\label{alpha1.2}
\end{figure}

Above calculations can be extended to the case of  the non-linear quenching of a term of the
Hamiltonian, e.g., with the variation of  the transverse field $h$ given  by $1-\rm{sgn} (t)(t/\tau)^{\alpha}$,
where sgn stands for the sign function of $t$. Although the probability of excitation 
is not exactly known, casting the Schr\"odinger equations which describe the time evolution of the two-level systems to a dimensionless form  \cite{sen08},
it has been argued  that $p_k$ should be a function of the dimensionless
combination of $k$ and $\tau$ given by
$p_k = G(k^2\tau^{2\alpha/(\alpha+1)})$, where $G$ is the scaling function. Considering only the contributions
from the low-energy modes for large $\tau$, one
finds the scaling
$D_{non-ad}(t)=\exp(-C N \delta^2 t^2/\tau^{\alpha/(\alpha+1)})$,
where $C$ is a number which also depends on $\alpha$. 
This is again in congruence with the scaling of $n$ for 
a non-linear quenching i.e., $n \sim \tau^{-\alpha/(\alpha+1)}$. 
This scaling has been numerically verified by directly integrating the Schr\"odinger equation
(see discussion around Eq.~(\ref{eq_gendecoh})) and results are presented in
Fig. \ref{alpha1.2}(a).

In order to extract the  exponent in a transparent way from the numerical data,
double logarithm of $D_{non-ad}$ is required which is numerically not possible
since $D$ is always less than unity. Hence, to calculate the exponent of $\tau$,
we introduce  a modified DF, $A(\tau,t)$ given by
$A(\tau,t)=-\log_{10} D_{non-ad}$. Fig.~\ref{alpha1.2}(a) clearly shows that  $\ln A(\tau, t)$ varies linearly with $\ln \tau$
and has a slope given by $-\alpha/(\alpha+1)$ with a fixed $t$, thus confirming the analytically predicted scaling
relation.


If the parameter $h$ is set equal to $2J_y$ and the interaction term $J_x$ is quenched as $t/\tau$, 
the spin chain (\ref{eq_xy1}) is driven across the quantum MCP  A at $J_x=J_y$ or $t=J_y \tau$ (see Fig.~\ref{fig_xy}); $n$ 
satisfies a scaling relation\cite{mukherjee07}  $n \sim \tau^{-1/6}$. This is not
in agreement with the KZ prediction and has been justified by asserting the existence
of quasi-critical points on the ferromagnetic
side of the MCP
\cite{mukherjee10}. 
 What would
happen if the environmental spin chain is driven across the  MCP? Choosing
appropriately the interaction $H_{SE}$ \ct{sharma12}, 
one finds
$\ln D_{non-ad}(t) \sim (t-J_y\tau)^2/\tau^{1/6}$~$ \sim (J_x-J_y) \tau^{11/6}$ 
(see Fig. \ref{alpha1.2}(b)). 

Our studies have so far  been limited to isolated quantum critical and multicritical points
and in all cases, the scaling of $\ln D_{non-ad}(t)$ (or $f(\tau)$)  with  $\tau$ is identical
to that of $n$, which is not necessarily given by the traditional KZ scaling.
Does  this scenario hold true in general? Below we highlight
a special situation where this connection between the scaling of $D_{non-ad}(t)$ and $n$, clearly breaks down.

Let us probe the scaling of $D_{non-ad}$ when the 
parameter $\ga$ of the environment
 (\ref{eq_xy1}) is  
quenched as $\ga=t/\tau$ so that the spin chain is
swept across the anisotropic critical point (for $|h| <1$) and the MCPs   along the gapless Ising transition lines for
$|h|=1$ (see Fig. \ref{fig_xy}). Note that here one rewrites Eq.~(\ref{eq_xy1}) in terms of $\gamma$ with $J_x+J_y=1$, and 
 modifies $H_{SE}$ to the form
$H_{SE}=-(\delta/2)\sum_i(\sigma_i^x\sigma_{i+1}^x-\sigma_i^y\sigma_{i+1}^y)\sigma_S^z$. This represents a CSM in which the 
CS
couples to the 
$XY$ spin chain  through the 
parameter $\gamma$. The coupling $\de$ therefore provides two channels of the
temporal evolution of the environmental ground state with anisotropy $\ga
+\de $ and $\ga-\de$, respectively. We recall that the problem was studied in Ref. \cite{divakaran08}
from the view point of defect generation.  For $|h|<1$,  $n \sim \tau^{-1/2}$, as
expected from KZ theory. For the DF,  one finds that  $\ln D_{non-ad} \sim t^2/\tau^{1/2} \sim \gamma^2 \tau^{3/2}$, which is also numerically verified
(see Fig. ~\ref{alpha1.2}(b)).  
Surprise emerges for $|h| =1$ where one finds $n \sim \tau^{-1/3}$, a scaling that can not be explained in terms of traditional KZ theory. 
Moreover, it was shown that
$p_k=e^{-2\pi \tau (1+\cos k)^2/\sin k} \sim e^{-\pi \tau k^3/2}$ for  $k \sim \pi$ when $h=1$.
Does this  imply a scaling  $\ln D_{non-ad}(t) \sim t^2/\tau^{1/3}$ for gapless quenching (see Fig.~1)?


To address
this question, we explore $h=1$ case in details.
Using an appropriate basis \ct{divakaran08}, one can recast the reduced $(2 \times 2)$ Hamiltonian $H_k(t)$
to the form
\begin{eqnarray}
H_k^\pm (t)&=& 2
 \left(
\begin{array}{cc}
  (\gamma \pm \delta) \sin k & h+\cos k  \\
h+\cos k & -(\gamma \pm \delta) \sin k   \\
\end{array}
 \right).
 \label{eq_matgamma}
\end{eqnarray}
Using Eq.~(\ref{eq_fknonad}) and noting  that $\Delta=4 \delta k$ we find that

\begin{eqnarray}
F_k=1-4(e^{-\pi \tau k^3/2}-e^{-\pi \tau k^3})\sin^2 (4 \delta k t)
\label{eq_gammafk}
\end{eqnarray}
for the modes close to $k=\pi$.  Assuming the limit $\de \to 0$ and  using mathematical steps identical to those employed
in deriving Eq.~(6) starting from Eq.~(5), we once again  find an exponential decay given by
\begin{eqnarray}
D_{non-ad}(t) \sim \exp\{-2^{14/3} N \delta^2 t^2/(3 \pi \tau)\}.
\label{eq_Dgamma}
\end{eqnarray}
We therefore find a clear deviation in the scaling of $D_{non-ad}$ (or $f(\tau) \sim \tau^{-1}$) from 
$n \sim \tau^{-1/3}$.
In the present case, the momentum dependence  of the term $\sin^2 (4 \delta k t)$
in Eq.~(\ref{eq_gammafk}) renders an additional $\tau^{-2/3}$ factor resulting to a $1/\tau$ scaling of $\ln D_{non-ad}$. 
This clearly presents a situation where there is no direct connection between $n$ and $D_{non-ad}$.
Substituting $t = \ga \tau$, one finds that $\ln D_{non-ad}  \sim -( 2^{14/3}N \delta^2 \gamma^2 \tau)/{3\pi}$;
this is numerically verified as shown in 
Fig.~\ref{alpha1.2}(b).
We note that the scaling (\ref{eq_Dgamma}) can also be reproduced analytically 
by solving the Schr\"odinger equation
with equivalent reduced Hamiltonian $H_k$ in Eq.~
 (\ref{eq_matgamma}).
In conclusion, we have found out the scaling
of the DF (or $f(\tau)$) of a qubit coupled to a quantum spin chain which is driven across
QCPs and quantum critical lines.
We show that the scaling of the DF is given by the scaling of $n$ for linear and non-linear quenching
through isolated critical points.
More importantly, our studies also reveal that
this scenario is not universally valid.

AD and UD acknowledge CSIR New Delhi, for financial support.

\end{document}